\documentclass[pdflatex,sn-nature,iicol]{sn-jnl}
\usepackage{graphicx,dcolumn,amsmath,amssymb,bm,xcolor}
\colorlet{RED}{red}
\begin{document}
\title{Beyond the CrI$_3$ Prototype: Rectangular Networks in Two-Dimensional Transition-Metal Fluorides}
\author*[1]{\fnm{Xin} \sur{Chen}}\email{xin.chen@iat.cn}
\author[2]{\fnm{Jin} \sur{Zou}}
\author[1]{\fnm{Zhen} \sur{Han}}
\author[1]{\fnm{Peng} \sur{Zhang}}
\author*[3]{\fnm{Xu} \sur{Cheng}}\email{xchengab@sdu.edu.cn}
\author*[2]{\fnm{Duo} \sur{Wang}}\email{duo.wang@mpu.edu.mo}
\affil[1]{Thermal Science Research Center, Shandong Institute of Advanced Technology, Jinan 250100, Shandong Province, People's Republic of China}

\affil[2]{Faculty of Applied Sciences, Macao Polytechnic University, Macao SAR, 999078, People's Republic of China}

\affil[3]{Institute for Advanced Technology, Shandong University, Jinan 250061, People's Republic of China}

\abstract{
The honeycomb lattice of CrI$_3$ has guided the search for two-dimensional transition-metal trihalides, yet this prototype overlooks a systematic preference for rectangular networks among fluorides. These networks have experimental counterparts in the layers of bulk post-perovskite NaFeF$_3$, NaNiF$_3$, and NaCoF$_3$. We compare honeycomb and rectangular monolayers across 32 compositions. Magnetic MACE phonon calculations indicate that 57 of the 64 structures are dynamically stable, including all eight rectangular fluorides. First-principles energies favor rectangular networks for seven of eight fluorides and honeycomb networks for 22 of 24 heavier halides. In rectangular FeF$_3$, an isolated upper valence band supports quasi-one-dimensional holes whose transport-distribution anisotropy exceeds 100 across a 0.20~eV window and peaks near 1200, while conduction-band electrons have nearly isotropic in-plane masses. Rectangular fluorides thus emerge as a distinct family of two-dimensional magnets, exemplified by the pronounced carrier-dependent transport anisotropy of r-FeF$_3$.
}
\maketitle

\section{Introduction}
\label{sec:introduction}
Since the experimental realization of monolayer CrI$_3$ in 2017, transition-metal trihalides have become a prominent platform for two-dimensional (2D) magnetism~\cite{Huang2017CrI3,Gibertini2019,Mak2019}. The honeycomb metal network of CrI$_3$, built from edge-sharing octahedra, has since provided a widely adopted structural starting point for exploring other MX$_3$ compositions~\cite{doi:10.1021/cm504242t,C5CP04835D,C5TC02840J,PhysRevB.98.144411,TOMAR2019165384,Wines2023}. Yet the same MX$_3$ stoichiometry need not imply the same low-energy connectivity: changing the transition metal or halogen changes the metal--halogen bond length and local octahedral geometry, and can therefore favor a different network. This issue is especially relevant for fluorides, where the metal--halogen bond lengths differ most strongly from the metal--iodine bonds in CrI$_3$. We therefore ask whether the CrI$_3$ honeycomb prototype still describes the relevant low-energy structures of 2D trihalides.

Our previous evolutionary structure search found that a rectangular CrF$_3$ monolayer, denoted r-CrF$_3$, is lower in energy than the CrI$_3$-type honeycomb structure~\cite{PhysRevB.111.155425}. In this structure, edge-sharing CrF$_6$ chains are connected through corner sharing, producing a network distinct from the CrI$_3$-type honeycomb framework. That work showed that this alternative connectivity is dynamically viable and electronically anisotropic~\cite{PhysRevB.111.155425}. The broader significance of the rectangular network nevertheless depends on two questions left open by the original CrF$_3$ prediction: whether it remains energetically competitive across related compositions, and whether it has a counterpart in experimentally established crystal chemistry.

A crystal-chemical counterpart to the rectangular connectivity appears in post-perovskite fluorides. Experimentally synthesized post-perovskite NaFeF$_3$ is a layered compound in which Fe--F sheets are separated by Na ions~\cite{Bernal2014NaFeF3,Lemoine2021NaFeF3}. Within each sheet, edge-sharing FeF$_6$ chains are connected through corner sharing, giving the same periodic Fe--F connectivity as the predicted r-FeF$_3$ monolayer. Related post-perovskite NaNiF$_3$ and NaCoF$_3$ phases show that this layered framework is not unique to Fe~\cite{LindsayScott2014Neutron}. These bulk phases do not constitute an experimental realization of an isolated monolayer, but they establish the rectangular fluoride framework as a known crystal-chemical motif and suggest precursor-based routes, such as desodiation and layer separation, for exploring neutral monolayers.

Large pretrained models are changing atomistic simulation through a
pretrain--adapt paradigm analogous to that in language and vision \cite{Brown2020GPT3,Dosovitskiy2021ViT,Bommasani2021FoundationModels,Li2026EdenGNN,Batatia2022MACE,Batatia2025MACEMP,Liao2026EquiformerV3}. In
machine-learning interatomic potentials, broad first-principles pretraining can
provide transferable representations of local atomic environments that are then
adapted to target systems with comparatively little additional DFT data. MACE
combines higher-order equivariant message passing with atomic-cluster
representations, and the MACE-MP family has been introduced as a foundation
model for atomistic materials chemistry~\cite{Batatia2022MACE,Batatia2025MACEMP}.
Recent magnetic extensions further incorporate atomic magnetic moments as
explicit degrees of freedom, making such potentials applicable to magnetic
materials~\cite{mMACE2026}. This capability is useful here because shallow
collective distortions and large commensurate supercells make exhaustive DFT
relaxation and phonon searches expensive. We therefore use pretrained magnetic
MACE as a transferable starting point and fine-tune it with DFT
data to resolve the soft modes in DFT calculations.

Here we compare honeycomb and rectangular networks for MX$_3$ monolayers with $M=$ Ti, V, Cr, Mn, Fe, Co, Ni, Cu and $X=$ F, Cl, Br, I, using first-principles energies for the lowest sampled magnetic state of each network. The rectangular network lies lower in energy for seven of the eight fluorides, with the two networks nearly degenerate in CoF$_3$, whereas 22 of the 24 heavier halides favor the honeycomb network. MACE-assisted relaxations and phonon calculations show that all 32
honeycomb candidates and all eight rectangular fluorides are stable without imaginary phonon frequencies.
Magnetic screening reveals diverse magnetic orders, including six
rectangular trihalides exhibiting type-IV collinear magnetism,
expanding the materials space for unconventional fully compensated
magnetism. Beyond the intrinsic flexural softness of two-dimensional
materials, rectangular fluoride networks accommodate cooperative
octahedral rotations, introducing additional soft modes that complicate
lattice-dynamical calculations. By fine-tuning a pretrained MACE potential,
we explore the shallow distortion landscape in large supercells and
identify a DFT-verified tilted minimum in r-FeF$_3$, providing a practical
route to investigating the vibrational and thermal properties of this
material family. In the resulting antiferromagnetic framework, directional Fe $d_{yz}$--F states support quasi-one-dimensional holes across a broad energy window, with a transport-distribution anisotropy much larger than that of monolayer black phosphorus~\cite{Rudenko2015BPModel}, whereas electrons at the predominantly Fe $d_{z^2}$ conduction edge have nearly equal in-plane effective masses. Together, these results move rectangular fluorides from an isolated CrF$_3$ prediction to a crystal-chemically grounded family of two-dimensional networks whose connectivity produces hole-transport anisotropy beyond that of black phosphorus.

\par

\section{Results and discussion}
\label{sec:results}
\subsection{Chemical trends in structural preference}
\label{sec:family_screen}

\begin{figure*}[htbp]
\centering
\makebox[\linewidth][c]{\includegraphics[width=17.18cm]{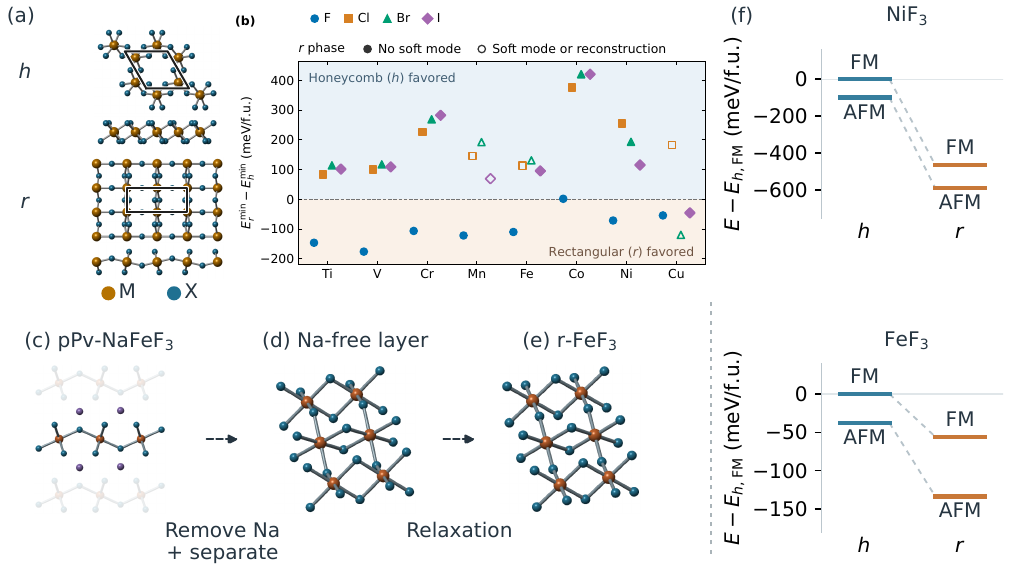}}
\caption{(a) Hexagonal and rectangular MX$_3$ monolayers. 
(b) Difference between the lowest sampled DFT energies of the two networks,
$E_r^{\min}-E_h^{\min}$, where the minimum is taken over the sampled spin
configurations. Filled symbols indicate rectangular candidates that retain their
r-network connectivity after magnetic-MACE relaxation and show no MACE phonon
branches below $-0.05$~THz. Open symbols indicate rectangular candidates that
either reconstruct during relaxation or show such modes. 
(c--e) Post-perovskite NaFeF$_3$, its Na-free $(010)$ layer, and relaxed
r-FeF$_3$. 
(f) Static DFT energies of the FM and selected AFM configurations of NiF$_3$
(top) and FeF$_3$ (bottom), referenced to the corresponding h-FM state of each
compound.}
\label{fig:family_screen}
\end{figure*}

Across the 32 MX$_3$ compositions ($M=$ Ti, V, Cr, Mn, Fe, Co, Ni, Cu
and $X=$ F, Cl, Br, I), the relative energies of the honeycomb and
rectangular networks depend strongly on the halogen
[Fig.~\ref{fig:family_screen}(b)]. Comparing the lowest static DFT
energies among the sampled magnetic configurations reveals a
rectangular preference for seven of the eight fluorides, extending
the earlier result for CrF$_3$~\cite{PhysRevB.111.155425}.
In CoF$_3$, both minima are nonmagnetic and differ by only
1.99~meV/f.u., leaving their energetic ordering unresolved at this
level. In contrast, 22 of the 24 chlorides, bromides, and iodides
favor the honeycomb network. The two exceptions are CuBr$_3$ and
CuI$_3$, for which the rectangular network is lower in energy by
119.5 and 45.5~meV/f.u., respectively.

To assess the dynamical viability of these structures, we performed
structural relaxations and phonon calculations using a pretrained
magnetic MACE potential. Of the 64 structural candidates, 57 retain
their network connectivity and exhibit no phonon branches below
$-0.05$~THz, comprising all 32 honeycomb and 25 rectangular
structures, including all eight rectangular fluorides.
The remaining seven rectangular candidates either reconstruct
upon relaxation or exhibit imaginary phonon frequencies. The details are give in the Supplementary.

Within the fluoride series, the rectangular preference is not tied to a single
choice of magnetic order in the representative cases examined here. Motivated by spin--lattice coupling in monolayer CrF$_3$~\cite{zheng2024spinlatticecouplinginducedrich}, we compare
FeF$_3$ and NiF$_3$ across FM and selected AFM states
[Fig.~\ref{fig:family_screen}(f)]. In both compounds, the rectangular network
remains lower in energy for the magnetic comparisons shown, and the same ordering
is recovered when each network is allowed to adopt its lowest sampled spin
configuration.

For FeF$_3$, the r-FM state is 55.8~meV/f.u. lower than h-FM, and r-AFM100 lies
96.0~meV/f.u. below the h-N\'eel state. Across the full set of sampled spin
configurations, the lowest rectangular state is AFM101, which is
109.7~meV/f.u. below h-N\'eel, the lowest sampled honeycomb state
[Fig.~\ref{fig:family_screen}(b) and Supplementary Table~S2].

NiF$_3$ shows the same trend. In the FM and selected AFM comparisons in
Fig.~\ref{fig:family_screen}(f), the rectangular network lies below the
corresponding honeycomb states. Allowing each network to adopt its lowest sampled
spin configuration gives
$E_r^{\min}-E_h^{\min}=-71.1$~meV/f.u. [Fig.~\ref{fig:family_screen}(b)]. The
honeycomb minimum has stripy order, whereas the rectangular minimum has a
three-up--one-down spin arrangement in the four-Ni magnetic cell
(Supplementary Table~S2).

\subsection{Crystal-chemical correspondence to post-perovskite fluorides}
\label{sec:structural_stability}
The Fe--F layers of bulk post-perovskite NaFeF$_3$ provide a
crystal-chemical counterpart to the network predicted for free-standing
r-FeF$_3$. The $Cmcm$ phase can be recovered at ambient conditions after
high-pressure synthesis~\cite{Bernal2014NaFeF3,Lemoine2021NaFeF3}. In this
structure, edge-sharing FeF$_6$ octahedra form chains along $[100]$, and
neighboring chains share corners along $[001]$, forming $(010)$ sheets separated
by Na ions. Neutron powder diffraction refinements of bulk post-perovskite
NaNiF$_3$ and NaCoF$_3$ establish the same layer connectivity for Ni and
Co~\cite{LindsayScott2014Neutron}.

In the structural comparison in Fig.~\ref{fig:family_screen}(c--e), removing
the Na sublattice and isolating one $(010)$ sheet leaves a stoichiometric
Fe$_2$F$_6$ framework. Each Fe retains six F neighbors, and the periodic Fe--F
connectivity matches that of relaxed r-FeF$_3$, with
$[001]_{\rm bulk}\parallel x$ and $[100]_{\rm bulk}\parallel y$. This
correspondence suggests a possible experimental test based on oxidative
desodiation,
$\mathrm{NaFeF_3}\rightarrow\mathrm{FeF_3}+\mathrm{Na}^{+}+e^{-}$, which would
raise the formal Fe valence from $+2$ to $+3$, followed by layer separation if
the fluoride framework survives.

\subsection{Magnetic diversity and type-IV collinear magnetism}
\label{sec:magnetic_classification}

The recent discovery of unconventional fully compensated collinear
magnetism has extended the classification of magnetic materials
beyond conventional ferromagnets and antiferromagnets~\cite{Smejkal2022Symmetry}.
Of particular interest in two dimensions is type-IV collinear
magnetism~\cite{Bai2025TypeIV,Tian2026TypeIV}, which combines nonrelativistic spin
degeneracy with the possibility of relativistic spin splitting
and time-reversal-symmetry-breaking responses despite vanishing
net magnetization. Unlike conventional antiferromagnets, the
opposite-spin sublattices in type-IV magnets are related by
out-of-plane mirror reflection $M_z$ or twofold rotation $C_{2z}$,
rather than inversion or translation. Despite recent advances
in their symmetry classification, material candidates remain
limited, particularly within chemically related structural families.

To examine the magnetic diversity of two-dimensional trihalides,
we classify the lowest-energy magnetic configurations obtained
for both honeycomb and rectangular networks.
The calculated magnetic orders span nonmagnetic, ferromagnetic,
ferrimagnetic, conventional antiferromagnetic, and unconventional
fully compensated collinear states
(Supplementary Table~S2).
Of particular interest, six rectangular trihalides exhibit type-IV
collinear magnetic symmetry: r-CrF$_3$, r-CuF$_3$, r-FeBr$_3$,
r-NiBr$_3$, r-FeI$_3$, and r-NiI$_3$.
The classification of r-CrF$_3$ is consistent with recent
spin-layer-group analysis~\cite{Du2026Stacking}, while the remaining
compounds extend this magnetic motif across different transition
metals and halogens.

These results also reveal a magnetic consequence of extending
structural searches beyond the CrI$_3$ prototype: restricting
the search to honeycomb lattices can exclude not only energetically
competitive structures but also entire sets of unconventional magnetic configurations.

\subsection{\texorpdfstring{r-FeF$_3$}{r-FeF3} as a representative rectangular fluoride}
\label{sec:rfeF3_representative}

\begin{figure*}[tbp]
\centering
\includegraphics[width=\linewidth]{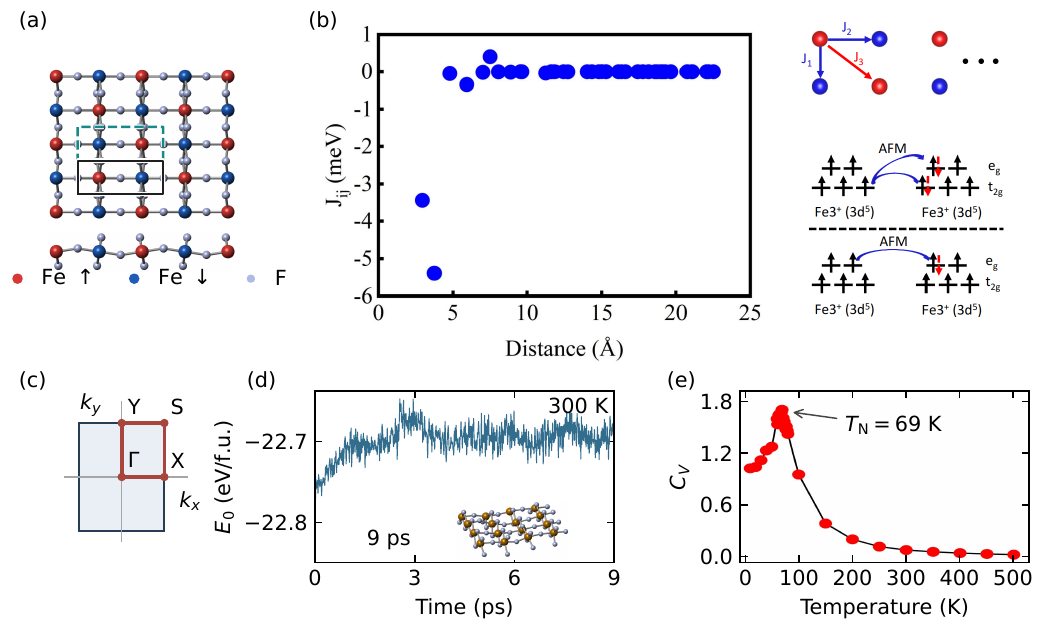}
\caption{Structure, magnetism, and finite-temperature behavior of AFM101
r-FeF$_3$. (a) VASP-relaxed tilted monolayer. \textcolor{black}{(b) RSPt exchange parameters $J_{ij}$ versus Fe--Fe distance (left), exchange paths $J_1$--$J_3$ (upper right), and schematic antiferromagnetic exchange channels between high-spin Fe$^{3+}$ ($3d^5$) ions (lower right). Negative $J_{ij}$ favors antiparallel alignment.} \textcolor{black}{(c) Magnetic Brillouin zone.} (d) AIMD trajectory at 300~K for the reference structure; the inset shows the final 9~ps. \textcolor{black}{(e) Magnetic specific heat $C_V$ from UppASD as a function of temperature.}}
\label{fig:Figure2}
\end{figure*}

Having established a fluoride-wide preference for rectangular networks, we use
r-FeF$_3$ as a representative member for the electronic and transport analysis
below. FeF$_3$ combines a sizable rectangular energy preference with the
crystal-chemical correspondence to post-perovskite NaFeF$_3$, making it a useful
case for connecting the structural trend to material-specific properties.

The VASP-relaxed tilted r-FeF$_3$ monolayer has lattice constants
$a=7.5022$~\AA{} and $b=5.9252$~\AA{} [Fig.~\ref{fig:Figure2}(a)]. Its nearly
coplanar Fe sites retain sixfold coordination, while F displacements tilt the
octahedra within the rectangular network. This tilted geometry and its magnetic
Brillouin zone [Fig.~\ref{fig:Figure2}(c)] define the structure and
reciprocal-space directions used for the unstrained electronic analysis.

\textcolor{black}{Among the 12 collinear configurations compared by DFT at the relaxed tilted geometry, AFM101 lies 6.78~meV/Fe below its nearest competitor. The exchange paths in Fig.~\ref{fig:Figure2}(b) explain this ordering. The two leading Fe--Fe couplings, $J_1=-3.437$ and $J_2=-5.389$~meV, both favor antiparallel moments along the sides of the Fe rectangle. The diagonal coupling, $J_3=-0.045$~meV, is too weak to compete appreciably with either interaction. Independent spin--orbit calculations place the easy axis out of plane ($E_x-E_z=0.1033$ and $E_y-E_z=0.0276$~meV/Fe). The magnetic specific heat peaks at approximately 69~K in the UppASD simulation [Fig.~\ref{fig:Figure2}(e)], giving an estimate of the N\'eel temperature.}

At 300~K, the structure retains sixfold Fe coordination and
its Fe--F bonding network during AIMD [Fig.~\ref{fig:Figure2}(d)], supporting
its finite-temperature structural robustness.

\subsection{Searching the shallow structural minimum of \texorpdfstring{r-FeF$_3$}{r-FeF3}}
\label{sec:mace_softmodes}

Soft phonon branches are a common numerical and physical complication in
two-dimensional materials, especially near $\Gamma$, where weak out-of-plane
restoring forces and residual stress can easily produce shallow imaginary
frequencies. The rectangular MX$_3$ network adds another soft degree of freedom:
edge-sharing chains connected by corner-sharing bridges can accommodate
cooperative octahedral rotations without immediately breaking the M--X
connectivity. Similar low-frequency distortions were already encountered in
r-CrF$_3$~\cite{PhysRevB.111.155425}. In r-FeF$_3$, the issue is more pronounced:
although the reference rectangular structure retains its Fe--F bonding network
during 300~K AIMD, its phonon spectrum contains soft modes near $\Gamma$ and at
finite wave vectors. AIMD alone therefore does not establish whether the
softening reflects a shallow local distortion, a larger-period structural
minimum, or an incipient reconstruction of the rectangular network. A reliable
phonon spectrum is therefore a prerequisite not only for assessing structural
stability, but also for any subsequent discussion of vibrational free energies,
thermal expansion, or phonon-mediated transport.

\begin{figure*}[tp]
\centering
\includegraphics[width=\linewidth]{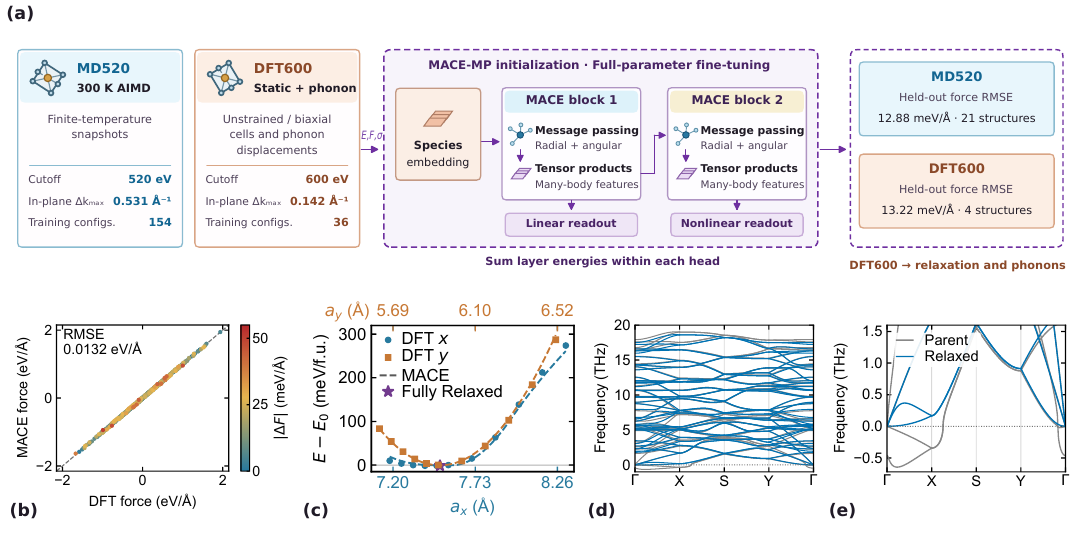}
\caption{MACE-assisted soft-mode relaxation of r-FeF$_3$.
(a) Joint fine-tuning of MACE-MP on AIMD and static DFT data with separate output heads. (b) Held-out DFT and MACE force components, colored by absolute error. (c) Relative energy versus lattice constant under
$x$ loading (blue, bottom axis) and $y$ loading (orange, top axis), with DFT
points, MACE curves, and the fully relaxed DFT tilted structure (star),
referenced to each method's unstrained structure. (d,e) Full and low-frequency
phonon spectra before and after MACE-assisted relaxation.}
\label{fig:mace_softmodes}
\end{figure*}

To identify the origin of these soft modes, we used a fine-tuned magnetic MACE
potential to explore the distortion landscape beyond the scale accessible to
direct DFT supercell relaxations [Fig.~\ref{fig:mace_softmodes}(a)]. Starting
from a pretrained MACE-MP potential~\cite{Batatia2022MACE,Batatia2025MACEMP},
we fine-tuned the model for AFM101 FeF$_3$ using DFT energies and forces from
AIMD snapshots, optimized structures, and phonon-displacement configurations,
thereby sampling both finite-temperature distortions and the local curvature of
the potential-energy surface.

The resulting model reproduces held-out DFT forces with errors of
0.0129--0.0132~eV/\AA{} and sub-meV/atom energy errors, depending on the data
setting [Fig.~\ref{fig:mace_softmodes}(a,b)]. We therefore used its
high-precision DFT head for the structural relaxations and phonon calculations
described below.

For the same $2\times2\times1$ phonon supercell, the fine-tuned MACE potential
reproduces the DFT softening of the reference structure, giving a minimum
frequency of $-1.21$~THz compared with $-1.23$~THz from DFT. In the larger MACE
phonon supercell, the reference structure remains soft, with a minimum frequency
of $-0.645$~THz and negative frozen-mode curvature. This agreement indicates
that the fine-tuned potential captures the relevant soft coordinate.

Relaxations from soft-mode-perturbed supercells of different sizes all converge
to the same shallow tilted minimum with the original magnetic-cell periodicity.
The distortion is a cooperative counter-rotation of neighboring FeF$_6$
octahedra that preserves sixfold Fe coordination and the rectangular Fe--F
connectivity. This tilted structure is lower
in energy by 0.425~meV/f.u. within the MACE description, and a small in-plane
lattice relaxation nearly removes the residual out-of-plane softening
[Fig.~\ref{fig:mace_softmodes}(d,e)]. Thus, the apparent soft instability
reflects a shallow octahedral-tilting coordinate coupled to residual in-plane
stress, rather than a tendency toward connectivity-breaking reconstruction or a
larger primitive cell.

DFT relaxation starting from the MACE-relaxed structure reaches the converge accuracy without any optimization, giving lattice constants $a=7.5022$~\AA{} and
$b=5.9252$~\AA{}, both within 1\% of the reference values. A matched-precision
DFT single-point comparison places this tilted structure 1.50~meV/f.u. below the
original reference. The asymmetric loading
curves further clarify the shallow nature of this distortion: extension from
compressed $x$ configurations initially unfolds the buckled Fe--F framework at
low energy cost, whereas further extension toward a straighter framework
increasingly involves bond stretching and a steeper energy rise.

Finite-supercell effects contribute to the residual small-cell softening. At the
same DFT-relaxed tilted geometry, increasing only the MACE phonon supercell from
$2\times2\times1$ to $4\times4\times1$ reduces the minimum frequency from
$-1.148$ to $-0.0034$~THz, whereas the corresponding $2\times2\times1$ DFT
calculation gives $-1.168$~THz. This suppression occurs even before applying
ASR, indicating that the residual imaginary branch is strongly supercell-size
dependent. We therefore use the DFT-verified tilted structure as the reference
geometry for the electronic and transport analysis below.

\begin{figure*}[tbp]
\centering
\includegraphics[width=\linewidth]{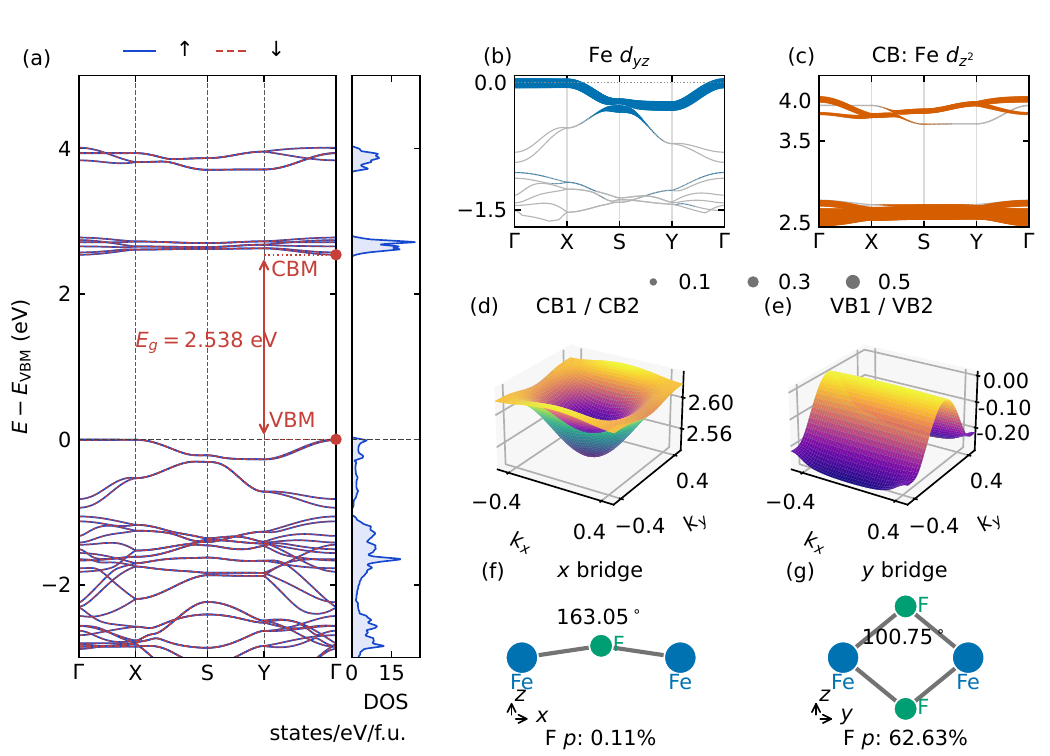}
\caption{Electronic structure and directional orbital channels of the relaxed
tilted geometry. (a) DFT band structure and density of states. (b,c) Orbital
projections of the valence Fe $d_{yz}$ and conduction Fe $d_{z^2}$ states, shown
with a common marker-area scale. (d,e) Wannier-interpolated energy surfaces of the conduction and valence bands, respectively. (f,g) Tilted-structure $x$- and
$y$-direction Fe--F--Fe bridges shown in $xz$ and $yz$ projections, with the
corresponding three-dimensional bond angles annotated. The F $p$ percentages are
summed over each bridge family within the top 0.10~eV of the valence band and
normalized to the total PAW-projected weight. Energies are referenced to the
VBM, and $k_x$ and $k_y$ are given in magnetic-cell reciprocal coordinates.
The orbital projections and band surfaces show one of the two degenerate spin channels. The DOS sums both spin
channels and uses a 0.02~eV Gaussian broadening.}
\label{fig:electronic_structure}
\end{figure*}

\begin{figure}[tbp]
\centering
\includegraphics[width=\linewidth]{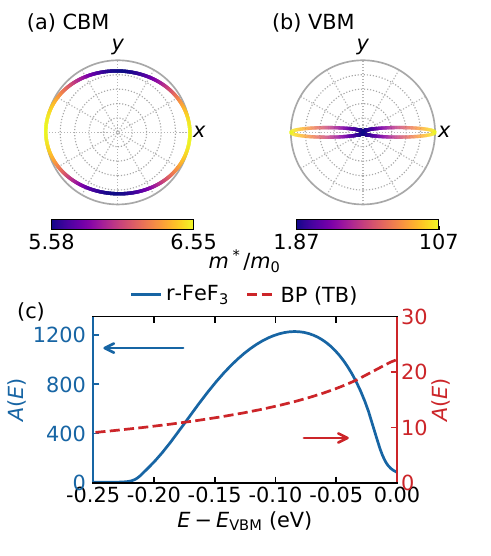}
\caption{Carrier-dependent anisotropy. (a,b) Directional electron and hole
effective masses. The radii are normalized to the maximum value for each carrier
type, with a radial grid spacing of 0.2, and the colors indicate $m^*/m_0$.
(c) Transport-distribution anisotropy for r-FeF$_3$ (blue solid line, left
axis) and monolayer black phosphorus (red dashed line, right axis). The two
vertical axes are scaled separately to compare the angular dependence rather
than the absolute magnitude.}
\label{fig:effective_mass}
\end{figure}

\subsection{Carrier-dependent band-edge anisotropy}
\label{sec:band_anisotropy}

Having established the relaxed rectangular framework and its magnetic order, we
next examine how this geometry shapes the band-edge electronic structure of
r-FeF$_3$. The AFM101 bands are spin degenerate without SOC. Both the
high-symmetry path and the two-dimensional DFT grid locate the VBM and CBM at
$\Gamma$, giving a direct gap of 2.538~eV
[Fig.~\ref{fig:electronic_structure}(a)]. The valence and conduction edges,
however, have markedly different in-plane dispersions. The highest valence band
disperses by only 5.18~meV along $\Gamma$--X but by 274.42~meV along
$\Gamma$--Y. The corresponding band-energy surface is therefore nearly flat
along $k_x$ and strongly dispersive along $k_y$, whereas the conduction-edge
surface has comparable curvatures along the two in-plane axes
[Fig.~\ref{fig:electronic_structure}(d,e)].

Directional effective masses are fitted to the Wannier bands along angles
$\theta$ measured from $x$ toward $y$. For
$E_s(k,\theta)=E_{0,s}+a_s(\theta)k^2$, the two spin-channel curvatures are
averaged:
\begin{equation}
m^*(\theta)=\frac{\hbar^2}{2|\bar a(\theta)|},\qquad
\bar a(\theta)=\frac{a_\uparrow(\theta)+a_\downarrow(\theta)}{2}.
\label{eq:massangle_main}
\end{equation}
The representative fit uses $|k|\leq0.06$~\AA$^{-1}$, and the extracted
anisotropy is insensitive to fitting windows between 0.02 and
0.06~\AA$^{-1}$. The DFT--Wannier band comparison and fitting-window tests are
shown in Supplementary Fig.~S4 and Supplementary Table~S3.

The corresponding hole masses are
$(m_x^h,m_y^h)=(107.21,1.87)m_0$, giving $m_x^h/m_y^h=57.44$, whereas the
electron masses are $(m_x^e,m_y^e)=(6.55,5.58)m_0$, giving
$m_x^e/m_y^e=1.17$ [Fig.~\ref{fig:effective_mass}(a,b)]. The long hole lobe
along $x$ represents a heavy mass and weak curvature, while the electron profile
is nearly circular. In a low-density rigid-band picture, the light
$y$-direction hole mass and heavy transverse mass favor strongly directional
hole transport, whereas the comparable electron masses favor a much more
isotropic electron response. Thus, electron and hole doping would probe
fundamentally different transport anisotropies in the same rectangular lattice.

\subsection{Broad-window anisotropy beyond the band-edge mass limit}
\label{sec:broad_transport}

To test whether the hole anisotropy is confined to the immediate vicinity of
$\Gamma$, we evaluated the sheet transport distribution from the Wannier bands,
\begin{equation}
\Xi_{ij}(E)=\frac{e^2}{A_{\rm cell}N_k}\sum_{n\mathbf{k}s}
v_{n\mathbf{k}s,i}v_{n\mathbf{k}s,j}\,
\delta(E-\epsilon_{n\mathbf{k}s}),
\label{eq:transport_distribution_main}
\end{equation}
where
$v_{n\mathbf{k}s,i}=\hbar^{-1}\partial\epsilon_{n\mathbf{k}s}/\partial k_i$,
$A_{\rm cell}$ is the in-plane cell area, and $N_k$ is the number of uniform
integration points. With a common scalar relaxation time, the conductivity is
obtained by thermal broadening of $\Xi_{ij}(E)$ around the chemical potential.
In the zero-temperature limit, the ratio
$A(E)=\Xi_{yy}(E)/\Xi_{xx}(E)$ therefore gives the rigid-band conductivity
anisotropy for a chemical potential placed at energy $E$.

Extreme hole directionality persists well beyond the local-curvature regime
[Fig.~\ref{fig:effective_mass}(c)]. The transport-distribution ratio reaches
$1.23\times10^3$ at 0.084~eV below the VBM, remains above $10^3$ over depths of
0.045--0.132~eV, and exceeds 100 down to 0.207~eV. This broad high-anisotropy
interval shows that the weak transverse response is not merely a consequence of
the band-edge mass at $\Gamma$, but persists across the upper valence manifold.

For black phosphorus, a benchmark for in-plane transport anisotropy~\cite{Xia2014BPAnisotropy,Qiao2014BlackPhosphorus}, the ten-hopping monolayer model fitted to $GW_0$ bands~\cite{Rudenko2015BPModel} gives an armchair-to-zigzag
transport-distribution ratio $\Xi_{\mathrm{AC}}/\Xi_{\mathrm{ZZ}}$ of
approximately 9--23 over the same 0--0.25~eV hole-energy interval. r-FeF$_3$
therefore shows much stronger hole directionality over most of the
0.05--0.20~eV window, although its ratio falls to about 6.10 at 0.25~eV below
the VBM.

These results establish that the large hole anisotropy is not a local
band-edge artifact, but a broad-window transport feature of the upper valence
manifold. r-FeF$_3$ therefore separates two transport regimes within the same
crystal: electron doping probes a nearly isotropic band edge, whereas hole
doping accesses a broad and highly directional valence manifold. This
carrier-selective anisotropy enables directional transport in a two-dimensional
magnetic semiconductor without changing the crystal orientation.

\begin{figure}[tbp]
\centering
\includegraphics[width=\linewidth]{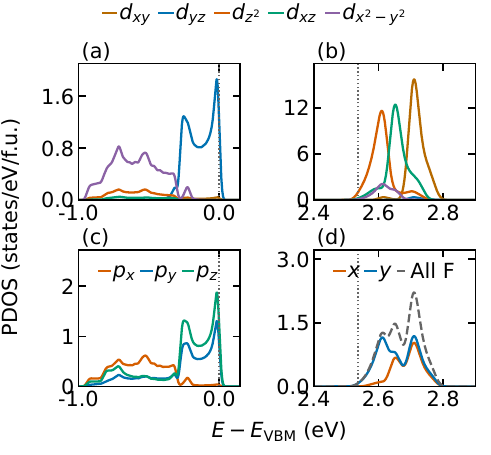}
\caption{Orbital-resolved DOS of the tilted structure. (a,b) Valence and conduction Fe $d$ states. (c) Valence F $p$ states. (d) Conduction F $p$ states on $x$ and $y$ bridges; dashed: total F. Energies reference the VBM; dotted lines mark band edges. Both spins are summed per formula unit, with 0.02~eV Gaussian broadening.}
\label{fig:edge_pdos}
\end{figure}

\subsection{Orbital origin of the strong hole anisotropy}
\label{sec:orbital_channels}

The lattice thermal conductivity of r-FeF$_3$, which depends on phonon dispersion and scattering as in other layered semiconductors~\cite{Zhang2016,C9CP01867K}, differs by only a factor of about two between the in-plane axes
(Supplementary Note V and Supplementary Fig.~S6). Thus, r-FeF$_3$ should not be
viewed as a van der Waals quasi-one-dimensional crystal in which strong lattice
anisotropy generically enforces quasi-one-dimensional behavior across different
properties. The much larger valence-band transport anisotropy instead points to
an electronic origin associated with specific orbital channels.

The difference between the band-edge mass anisotropy and the larger
transport-distribution anisotropy further indicates that the hole response is
not determined by the local curvature at $\Gamma$ alone. Because
Eq.~\eqref{eq:transport_distribution_main} weights squared velocities over the
full constant-energy manifold, a broad high-anisotropy window requires a
directional electronic channel that persists across the upper valence manifold.

This channel is provided by the Fe $d_{yz}$ and $y$-bridge F $p$ states
[Fig.~\ref{fig:electronic_structure}(b) and Fig.~\ref{fig:edge_pdos}(a,c)].
At $\Gamma$, the highest valence state contains substantial Fe $d_{yz}$,
F $p_y$, and F $p_z$ character, with the ligand weight residing mainly on the
F sites that connect Fe neighbors along $y$. In the global Cartesian frame, the
Fe $d_{yz}$ orbital has a nodal plane at $y=0$, which suppresses coupling to
the $x$-bridge ligands. The off-axis $y$ bridges instead allow strong
$d_{yz}$--$p_y/p_z$ hybridization, producing a predominantly $y$-directed Fe--F
channel consistent with the weak $\Gamma$--X dispersion.

This orbital selectivity persists over the hole-energy window. Within
0.10~eV below the VBM, Fe $d_{yz}$ and $y$-bridge F $p$ states dominate the
summed PAW weight, whereas the $x$-bridge F $p$ contribution remains negligible.
The same hierarchy is retained over 0.25~eV below the VBM, providing the
microscopic basis for the broad quasi-one-dimensional hole response.

The conduction edge has a different orbital character
[Fig.~\ref{fig:electronic_structure}(c) and Fig.~\ref{fig:edge_pdos}(b,d)].
The CBM is dominated by Fe $d_{z^2}$, with smaller contributions from other
Fe $d$ orbitals and F $p$ states. Unlike $d_{yz}$, the $d_{z^2}$ component has
a much less selective in-plane angular distribution, explaining why the electron
effective masses are nearly isotropic at the band edge.

The tilted structure also shows that the hole channel is not selected simply by
the largest Fe--F--Fe bond angle [Fig.~\ref{fig:electronic_structure}(f,g)].
Although the single $x$ bridge is more linear than the two $y$ bridges, its
F $p$ contribution to the upper valence window remains small. The strong hole
anisotropy is therefore controlled primarily by orbital symmetry and bridge
selectivity, rather than by structural anisotropy alone.

Supplementary Note II and Fig.~S3 examine how uniaxial strain tunes the bands of r-FeF$_3$, extending similar studies of puckered semiconductors~\cite{Fei2014StrainBP,Rodin2014StrainBP,PhysRevB.91.085423,doi:10.1002/anie.201411246} and magnetic layers~\cite{10.1063/5.0147450,Ma2026BipolarValley,Zhang2025RuCl2Strain}.

\subsection{Overview of fluoride band structures}

\begin{figure*}[tbp]
\centering
\includegraphics[width=\linewidth]{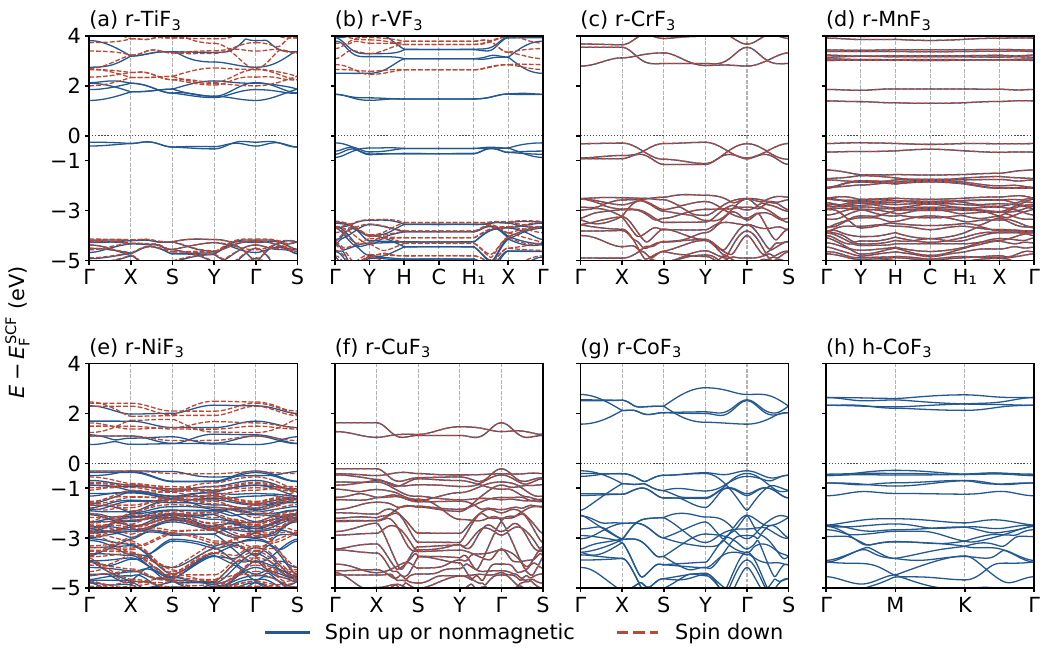}
\caption{Electronic bands of (a--g) rectangular Ti, V, Cr, Mn, Ni, Cu, and Co trifluorides and (h) honeycomb CoF$_3$. Each panel uses its own SCF Fermi level as zero. Blue solid and red dashed lines denote spin up and spin down, respectively. Nonmagnetic bands are shown in blue.}
\label{fig:fluoride_family_bands}
\end{figure*}

Figure~\ref{fig:fluoride_family_bands} gives an initial survey of the electronic bands of the other rectangular fluorides, with honeycomb CoF$_3$ included for comparison. Along the sampled paths, the upper valence bands of r-CrF$_3$, r-NiF$_3$, and r-CuF$_3$ are flatter along $\Gamma$--X than along $\Gamma$--Y. These compounds are natural candidates for testing how far the directional hole response established for r-FeF$_3$ extends across the family. The two nearly degenerate, nonmagnetic CoF$_3$ polymorphs offer a complementary opportunity to examine how octahedral connectivity affects the band edges at fixed composition. Supplementary Figs.~S11--S14 show the band structures of all h and r phases that passed the MACE phonon stability check.

\section{Conclusion}
\label{sec:conclusions}

Our results show that the CrI$_3$ honeycomb lattice is not a universal prototype
for two-dimensional transition-metal trihalides. In the chromium series, Cl, Br
and I favor the honeycomb network, whereas CrF$_3$ was previously found to favor
a rectangular network~\cite{PhysRevB.111.155425}. The same rectangular
preference recurs across the newly screened transition-metal fluorides, while
most heavier halides remain closer to the honeycomb prototype. Thus, fluorides
define a composition regime in which rectangular and honeycomb octahedral
networks must be treated as competing structural motifs before magnetism or
electronic structure is interpreted.

Our dynamical screening identifies 57 stable monolayer structures
out of 64 candidates, comprising 32 honeycomb and 25 rectangular
networks. Magnetic screening reveals diverse magnetic orders,
including six rectangular trihalides with type-IV collinear magnetism.
These results demonstrate that expanding the structural search space
beyond the CrI$_3$ prototype substantially enriches the materials
landscape for unconventional fully compensated collinear magnetism.

The r-FeF$_3$ case also highlights a practical challenge for this structural
branch. The free monolayer lies on a shallow octahedral-tilting landscape, where
small-cell DFT relaxations may miss the relevant minimum, while exhaustive
large-supercell phonon searches remain computationally prohibitive. A fine-tuned
magnetic MACE potential offers an efficient way to explore such soft distortion
landscapes and to separate shallow tilting coordinates from larger-period
instabilities or connectivity-breaking reconstructions. After DFT validation,
the resulting tilted structure provides a reliable reference for phonon-related
properties, including structural stability, vibrational free energies, thermal
expansion and lattice thermal transport.

The low-energy octahedral network favored by most fluoride monolayers in our
comparison is already realized in the layers of known post-perovskite compounds.
This connection recasts rectangular CrF$_3$ as an indication of a broader
structural branch of two-dimensional fluorides and reveals why the CrI$_3$ honeycomb prototype cannot be chemically extrapolated
across the full halide series. The fluoride series cannot be
treated simply as lighter-halogen copies of the same lattice: the preferred
octahedral connectivity is itself part of what changes across the series. In
r-FeF$_3$, this alternative connectivity supports sharply different carrier
motion because its valence and conduction states access different orbital
channels. A single connected layer can consequently support quasi-one-dimensional
holes alongside a nearly isotropic electron band edge. Together, these results
bring an experimentally established fluoride architecture into the structural
basis of two-dimensional magnetism and identify carrier-selective electronic
motion as a concrete consequence of that expanded picture.

\section{Methods}
\label{sec:methods}

First-principles calculations used VASP~\cite{002230939500355X,PhysRevB.54.11169},
projector augmented-wave potentials~\cite{PhysRevB.50.17953,PhysRevB.59.1758}
and the local spin-density approximation with the Dudarev $+U$
correction~\cite{Dudarev1998}, with
$U_{\rm eff}=3$ eV for metal $3d$ orbitals. Final calculations used plane-wave cutoffs
$\geq520$ eV, $\Gamma$-centered meshes with in-plane spacing
$\leq2\pi\times0.025$ \AA$^{-1}$ and electronic convergence
$\leq10^{-7}$ eV. FeF$_3$ relaxations used a force threshold of
0.001 eV/\AA{} at fixed vacuum spacing. Network comparisons used the lowest
sampled collinear or nonmagnetic static DFT energies, excluding entropy
contributions, per formula unit. Unstrained electronic and magnetic
calculations used the tilted FeF$_3$ structure; strain calculations used the nearly planar
reference, relaxing internal coordinates and the transverse lattice parameter.

Family screening used pretrained magnetic MACE to relax atomic positions and in-plane
cells at fixed DFT local moments, followed by $4\times4\times1$ finite-displacement phonon
calculations. For AFM101 FeF$_3$, MACE-MP~\cite{Batatia2022MACE,Batatia2025MACEMP}
was fine-tuned on AIMD and static DFT data; relaxations and phonons used its
DFT600 head (Sec.~\ref{sec:mace_softmodes}). DFT phonons used
PHONOPY~\cite{phonopy}, $2\times2\times1$ magnetic supercells and paired
$\pm0.01$ \AA{} displacements, with reference-force subtraction,
force-constant symmetrization and the translational acoustic sum
rule~\cite{Lin2022Invariance}. AIMD ran for 9 ps at 300 K.
Lattice thermal conductivity was calculated in the three-phonon relaxation-time approximation~\cite{Togo2015Phono3py}, using the outermost F--F layer thickness plus twice the F van der Waals radius~\cite{doi:10.1021/j100785a001} for normalization.

{\color{black}The Fe--Fe exchange parameters $J_{ij}$ were calculated for the AFM101 reference state using the magnetic force theorem in the full-potential linear muffin-tin orbital (FP-LMTO) code RSPt~\cite{wills2010full,kvashnin2015exchange,PhysRevB.111.155425}. The effective spin Hamiltonian is
\begin{equation}
\mathcal{H}=-\sum_{i\ne j}J_{ij}\,\mathbf{S}_i\cdot\mathbf{S}_j-\sum_i K_i(S_i^z)^2,
\end{equation}
where $\mathbf{S}_i$ is a unit vector along the local Fe moment at site $i$, and $K_i$ is the single-ion anisotropy constant for the out-of-plane $z$ axis. Negative $J_{ij}$ favors antiparallel moments. The magnetocrystalline anisotropy was evaluated in VASP using the force theorem with a common collinear charge density. The N\'eel temperature was estimated from the peak in the magnetic specific heat obtained by classical Monte Carlo simulations with UppASD~\cite{Skubic2008UppASD}.}

Wannier90~\cite{MOSTOFI20142309} provided a 56-orbital Fe $d$--F $p$
Hamiltonian for each spin channel, validated against independent DFT
eigenvalues. Effective masses and transport distributions used Wannier bands
and analytic velocities. Conductivity anisotropies were interpreted within
the rigid-band and common scalar relaxation-time approximations.
The black-phosphorus comparison used the same transport observable and
energy resolution. Validation and convergence tests are given in Supplementary
Information, Secs.~III and IV.


\section*{Acknowledgements}
This work is supported by the National Natural Science Foundation of China (Grant No.~12604277) and Shandong Provincial Natural Science Foundation (ZR2024QA040). We would like to thank Shandong Institute of Advanced Technology, National Supercomputing Center (Shuguang), and the ChinaHPC cloud computing platform (www.chinahpc.cn) for providing computational resources. Duo Wang acknowledges financial support from the Science and Technology Development Fund of Macao SAR (Nos.~0062/2023/ITP2 and 0016/2025/RIA1) and Macao Polytechnic University (Grant No.~RP/FCA-03/2023).

\section*{Competing interests}
The authors declare no competing financial interest.

\section*{Data availability}
The data that support the findings of this study are available from the
corresponding author upon reasonable request.

\bibliography{ref}

\end{document}